\pdfoutput=1
\pdfoutput=1
\pdfoutput=1
\pdfoutput=1
\pdfoutput=1
\documentclass[fleqn]
{report}

\usepackage[T1]{fontenc}
\usepackage{amsfonts,amssymb,amsmath} 
\usepackage{bm}
\usepackage[all]{xy}
\usepackage{graphics,graphicx,epsfig}
\usepackage{color}
\usepackage{fancyhdr}
\fancypagestyle{plain}{%
    \fancyhead{}                      

}
\usepackage [ english] { babel }
\usepackage{makeidx}
\makeindex
\usepackage{xr-hyper}
\usepackage[unicode,pagebackref]{hyperref}
\hypersetup{pdftex}

\newcommand{\beq}{\begin{equation}}
\newcommand{\eeq}{\end{equation}}
\newcommand{\beqa}{\begin{eqnarray}}
\newcommand{\eeqa}{\end{eqnarray}}

 \title{On basis images for the digital image representation
 }
      \author{Gorbachev V.N.,  Denisov L.A.,  Kaynarova E.M., Metelev I.K. 
       \footnote{High School of Printing and Mediatechnology of St.~Petersburg State University of Industrial Technology and Design, St.~Petersburg, Russia},
Yakovleva E.S. \footnote{St.~Petersburg State University, St.~Petersburg, Russia}  
                }
\begin{document}
\maketitle
\begin{abstract}
Digital array orthogonal transformations  that can be presented as a decomposition over basis items or basis images  are considered.
The orthogonal transform provides digital data scattering, a process of  pixel energy
redistributing, that is illustrated with the help of basis images. Data scattering plays
important role for applications as image coding and watermarking.
We established a simple quantum analogues of basis images. They are representations of  quantum operators that  describe transition of single particle between its states.
\\
Considering basis images as items of a matrix,
we introduced a block matrix  that is suitable
for orthogonal transforms of multi-dimensional  arrays such as block vector, components of which are matrices. We present an orthogonal transform that produces correlation between arrays. Due to correlation new feature of data scattering was found. A presented detection algorithm is an example of how it can be used in frequency domain watermarking.

 \end{abstract}

\tableofcontents
\section{Introduction}
A digital image has various  representations and   some of them are required by applications. Many useful representations
are produced by orthogonal transforms that are  powerful  tools  of image processing.  Well known  examples are JPEG  and JPEG2000 lossy compression formats based on DCT (Discrete Cosine Transform) and DWT (Discrete Wavelet  Transform).  For  the image compression problem block based DCT and DWT techniques are developed
\cite{blockDCT}  and generalized to non-separable transforms   \cite{nsepDCT}.
\\
Orthogonal transform  produces scattering of digital data, a process  that redistributes   pixel energy of transformed image. It is useful for protection the hiding  data in steganography, when a message is embedded into  image.  The hidden data  is scattered  among  all digital cover image  and   becomes  more robust  to  lossy data compression and  some statistical  attacks \cite{Graystg}.

The orthogonal transform of images may be considered as a decomposition over  matrices known as basis matrices \cite{PrattW}. Being some kind of grayscale images, the basis matrices look attractive and they are often reproduced by  textbooks \cite{Salomon}.
We will also call these  matrices  basis images.

In this paper we study basis images.
We focus on the  following questions:  color and wavelet  basis images,
orthogonal transform by matrix
of basis images and their quantum analogues.
For color images the solution  is  directly achieved by considering three-dimensional orthogonal transform  but for wavelets the solution is not so simple. The reason
is that  in practice, DWT is calculated  by
algorithms using signal processing techniques instead of orthogonal transforms.  Nevertheless these algorithms can be used to calculate wavelet basis images.  So it was found
for various wavelets  that  the basis has a block structure similar to   DWT coefficients  \cite{FRUCt17}.
\\
 Basis images may be considered as items of a matrix. We introduced such a  matrix,
it is orthogonal and suitable for transforms
of multi-dimensional arrays such as a  block vectors consisting of matrices.
In this case there is a large number of degrees of freedom that may be correlated by the transformation. The correlation results in new features of the orthogonal transform   for data scattering.
\\
Indeed, in the standard image transform  a given image pixel maps into all pixels of the transformed image.
To retrieve it back all the transformed image pixels
are required rather than one of them.
A new feature is that retrieval can be made from a single  pixel only due to correlation between arrays.
A detection algorithm illustrates how this feature may be used for the frequency domain watermarking.
\\
\\
The paper is organized as follows.
First, the  orthogonal transform and the  data scattering  and basis  images are considered.
Then a matrix consisting of  basis images and the orthogonal transform of  multi-dimensional  arrays are introduced. Next, an example of the  scheme for frequency domain watermarking is presented.

\section{The orthogonal transform}
The orthogonal transform can scatter digital data.
\\
\\
\textbf{The orthogonal matrix}
A square matrix $U$ of real  items is   orthogonal  if \cite{StrengAl}
\begin{eqnarray}
\label{001}
&&UU^{T}=1 \   \text{or} \ U^{T}U=1.
\end{eqnarray}
Columns of this matrix
 $u_{m}$ and rows $u^{T}_{n}$
are orthonormal vectors
\begin{eqnarray}
\label{0001}
\nonumber
&& \langle u_{m} u_{n}\rangle=\delta_{mn},\\
&&\langle u^{T}_{m} u^{T}_{n}\rangle=\delta_{mn},
 \end{eqnarray}
where $\langle x y\rangle$ denotes  scalar  product of two vectors.
\\
\\
\textbf{Scattering.} We will study  data scattering that can be  illustrated  by orthogonal transform of  vectors.
\\
Let us assume  that $f=\{f_{k}\}$, $k=1,\dots N$
is a vector
and $U$  is  an
$N\times N$ orthogonal matrix. Taking into account  that $f=UU^{T}f$, we find orthogonal transform of vector $f$
\begin{eqnarray}
\label{f102}
\nonumber
&&f=Ug.\\
&&g=U^{T}f,
\end{eqnarray}
where  vector $g=\{g_{p}\}$, $p=1,\dots N$
is often called a representation of $f$.
In matrix form  these equations look as follows
 \begin{eqnarray}
\label{f103}
\nonumber
&&f_{k}=\sum_{p} U_{kp}g_{p},\\
&&g_{p}=\sum_{k} U_{kp} f_{k}.
\end{eqnarray}
As a result two points concerning    data scattering can be made.
\begin{enumerate}
\item Every item of $f$ transforms into all items of $g$ with the  weight  $U_{kp}f_{k}$, where $p=1,2,\dots.$
\item To get $f_{k}$,  we need to know all items of $g$.
\end{enumerate}
Let us assume that the data are hidden in $f_{k}$, for example, by steganography and is  distributed among all   the  digital space of $g$ by an orthogonal transform.
The data can be extracted,  we need
all  space of $g$ in spite of every point
the data have. Formally the problem is to find $f_{k}$ for given $g_{p}$  and the orthogonal transform.
\\
Here and later we will consider data scattering as mapping
 \begin{eqnarray}
\label{fa104}
   f_{k} \leftrightarrows \{g_{1},g_{2},\dots \}.
   \end{eqnarray}
It is clear  that  due to symmetry, the vector   $f$ may be replaced with $g$. The considered  features are true for  the orthogonal transforms  of matrices and other  multi-dimensional arrays.

Data scattering can be directly demonstrated by  the orthogonal transform of a set of basic vectors. Let us consider a set of vectors and each of them
has a nonzero component
$e_{k}=\{\delta_{kn}\}$, $k,n=1,\dots, N$.
The vectors are known  to be unit vectors and form a standard basis \cite{StrengAl}
  \begin{eqnarray*}
e_{1}=\begin{bmatrix}
    1 \\
    0\\
    \vdots\\
    0
      \end{bmatrix},
      \ \
 e_{2}=\begin{bmatrix}
    0 \\
    1\\
    \vdots\\
    0
      \end{bmatrix},
      \ \
 \dots
 e_{N}=\begin{bmatrix}
    0 \\
   0\\
    \vdots\\
    1
      \end{bmatrix}.
\end{eqnarray*}
An orthogonal matrix $U$ transforms the
standard basis into another  basis consisting of the columns of $U$
\begin{eqnarray}
\label{f1a0110}
u_{k}=Ue_{k}.
\end{eqnarray}
This equation shows  that a single nonzero
item of $e_{k}$ distributes among a column $u_{k}$
\begin{eqnarray*}
e_{k}=
\begin{bmatrix}
    0 \\
        \vdots\\
        1\\
    \vdots\\
    0
      \end{bmatrix}
      \leftrightarrows
       \begin{bmatrix}
    u_{1k} \\
        \vdots\\
        u_{kk}\\
    \vdots\\
    u_{N}
      \end{bmatrix}
      =u_{k}.
 \end{eqnarray*}
Since  the column has at least two non zero items this transformation can be considered as scattering.

Scattering may results in  energy
concentration, a process that is important for applications.
The array energy, defined as
the sum of all components squared, is preserved under orthogonal transforms.
Due to scattering, the energy can
be distributed into a small amount of
components, that is a  base of coding in the image  compression field. It depends on the orthogonal matrix, regardless of whether the energy would  be concentrated or not.
It is known that DCT,  WHT (Walsh Hadamard Transform) and  KLT (Karhunen Loeve Transform) can   concentrate the image energy, if image is not random,
but  DST  (Discrete Sine Transform) can't do it  \cite{Salomon}.

\section{Basis images}

The orthogonal transform of a matrix and
three-dimensional array provides decomposition  over the grayscale and color basis images.
\\
\\
\textbf{Representation of matrix}.
Let $F=\{F_{mn}\}$ be a real rectangular $M\times N$ matrix, that corresponds to
a grayscale image. We introduce two orthogonal matrices $U=\{U_{mn}\}$ and $V=\{V_{pk}\}$
of $M\times M$ and $N\times N$. Then taking into account, that
$F=UU^{T} FVV^{T}$, we find
\begin{eqnarray}
\label{f104}
\nonumber
&&F=UGV^{T},\\
&&G=U^{T}FV. 
\end{eqnarray}
where $G$ is a  $M\times N$ matrix.
\\
Let us assume that $F$ is  an image in a spatial domain (that is the image as we see it). Matrix $G$ is usually called a frequency representation of $F$ or an  image in frequency domain.
The  frequency domain image
 may look senseless, however the orthogonal transform is reversible  and the original image can always be retrieved.

Using the matrix form of (\ref{f104}), for example,
\begin{eqnarray*}
&&F_{xy}=\sum_{kp}U_{xk}G_{kp}(V^{T})_{py}=
\sum_{kp}(u_{k}\otimes v_{p})_{xy}G_{kp},
\end{eqnarray*}
we get   a decomposition
over tensor products of rows and columns of the matrices  $U$ and $V$. Here and later we assume $U=V$ and $M=N$  that is a more interesting case.
Then the decomposition produced by the
orthogonal transformation  takes the form
\begin{eqnarray}
\label{f105}
\nonumber
&&F=\sum_{k,p} (u_{k}\otimes u_{p}) G_{kp} ,\\
&&G=\sum_{x,y}(u^{T}_{x}\otimes u^{T}_{y})F_{xy}.
\end{eqnarray}
We introduce the  matrices
\begin{eqnarray}
&&a_{kp}=u_{k}\otimes u_{p},\\
\nonumber
&&d_{xy}=u^{T}_{x}\otimes u^{T}_{y},
\end{eqnarray}
that  we  call  basis images.
There are $N^{2}$ basis images of size $N\times N$,  every  image pixel is a
product of two  items of the orthogonal matrix $U$
 \begin{eqnarray*}
&&a_{kp}(x,y)=U_{xk}U_{yp}.
\end{eqnarray*}
\textbf{Color basis images.}
A color RGB  image is a three-dimensional array and  similar to matrices it provides a decomposition over basis images.
\\
Let $T=\{T_{mnz}\}$ be a three-dimensional  array of $M\times N\times Z$.
The orthogonal transform of $T$ can be achieved by three orthogonal matrices $U$, $V$ and $W$. The matrices have size  $M\times M$, $N\times N$ and  $Z\times Z$  respectively.  Similarly  to (\ref{f105}) the array $T$ can be presented  as follows
\begin{eqnarray}
\label{b106}
T=\sum_{kps}(u_{k}\otimes v_{p}\otimes w_{s})\tau_{kps},
\end{eqnarray}
where $w_{s}$, $s=1,\dots, Z$ is a  column of  the matrix $W$. Tensor products
\begin{eqnarray*}
\label{e2018}
t_{kps}=u_{k}\otimes  v_{p}\otimes w_{s}=a_{kp}\otimes w_{s}
\end{eqnarray*}
produce a basis, the basis items  are
 \begin{eqnarray}
\label{e02018}
t_{kps}(m,n,q)=a_{kp}(m,n)w_{q}.
\end{eqnarray}
In general a  three-dimensional  array  can  not be a color image. The color  RGB image is  described by  three matrices $R$, $G$
and $B$ of equal dimensions, say $M\times N$. Matrices  are concatenated in a
$N\times N\times 3$ array
\begin{eqnarray*}
C=cat(3,R,G,B),
\end{eqnarray*}
where  $cat$ is concatenation.
Here we use notation of  MATLAB,
it means  that $C_{mn1}=R_{mn}$,
$C_{mn2}=G_{mn}$ and $C_{mn3}=B_{mn}$.

Let us assume  that in (\ref{b106}) $W$ is a  matrix of size $3\times 3$ and  introduce  color basis images
\begin{eqnarray*}
t_{kps}=cat(3, r_{kps},g_{kps},b_{kps}).
\end{eqnarray*}
Using (\ref{e02018})  we find the color channels
 $r_{kps}=a_{kp}w_{1s}$, $g_{kps}=a_{kp}w_{2s}$ and  $b_{kps}=a_{kp}w_{3s}$.  Full basis has $M\cdot N$ color items.  As a  result we get the  decomposition of  RGB images over basis color images
\begin{eqnarray*}
C =\sum_{kps}cat(3,r_{kps},g_{kps},b_{kps}) \tau_{kps}.
\end{eqnarray*}

\section{Properties of basis images}
Being tensor products of  orthogonal matrix columns and rows  the basis images have properties  that follow from orthogonality, and they have a simple  analogue came  from quantum mechanics.
\\
\\
\textbf{ Properties.}
Now let us consider the basis images
$a_{kp}$, if  $U=V$,  properties of $d_{xy}$ are the same.
\begin{enumerate}
\item  The matrix product of two basis images is another basis image
\begin{eqnarray*}
&&a_{kp}\cdot a_{mn}=a_{kn}\delta_{pm}.
\end{eqnarray*}
\item The scalar product
\begin{eqnarray}
\label{b3000}
&&\langle a_{kp}, a_{mn}\rangle =\delta_{km}\delta_{pn},
\end{eqnarray}
where the scalar product of matrices is
$\langle A,B\rangle=\sum_{mn}A_{mn}B_{mn}$.
\item The sum of  diagonal elements, trace
\begin{eqnarray}
\nonumber
\label{b3003}
&& \sum_{k}a_{kk}= 1, \\
&&\sum_{x}a_{kp}(x,x)=\delta_{kp}.
\end{eqnarray}
It follows that $\sum_{k} a_{kk}(xy)=\delta_{xy}$.
\end{enumerate}
Analysing
 these properties
we came to the conclusion that
basis images are orthonormal.
This observation allows us to consider
the orthogonal transform  (\ref{f105})   as a standard decomposition  over
the orthonormal basis. Is is obvious that  the first equation  takes the form
\begin{eqnarray}
\label{f305}
&&F=\sum_{k,p} a_{kp} G_{kp},
\end{eqnarray}
where $ G_{kp}= \langle F,a_{kp}\rangle.$
\\
\\
\textbf{Generation of basis images.}
There are at least two ways  to get basis images.
The first is to use its definitions. In this case
the orthogonal matrix has to be given.
The second way follows from orthogonal transform of the basis images.
\\
Let us focus on the second approach.
Let  $F=a_{kp}$  be in equation
(\ref{f305}). Then we find the basis image representation of the form  $G_{k,p} =\delta_{ka}\delta_{pb}$.
It means that the matrix $G$ has one nonzero pixel, it is equal to   1 and its position is $(a,b)$. So, the orthogonal transform of a basis image is a binary matrix of unit brightness. We denote such unit matrix as
\begin{eqnarray}
\label{b3010}
&& e_{ab}=\{\delta_{ka}\delta_{pb}\},
 \end{eqnarray}
 where  $k,p=1,\dots,N$.
Then the next relations are true
\begin{eqnarray}
\label{b3012}
&&a_{ab}=Ue_{ab} U^{T},\\
\nonumber
&& d_{ab}= U^{T}e_{ab}U.
\end{eqnarray}
These equations are two-dimensional analogue of (\ref{f1a0110}) and they have a simple meaning.
So  together with the unit vectors $e_{k}$ the unit matrices $e_{ab}$ form a standard basis and the  orthogonal transform of the basis is a set of basis images $a_{ab}$.
\\
Indeed, with the help of  the  standard basis any matrix  can be presented  in the following form
\begin{eqnarray*}
G=\sum_{kp}G_{kp}e_{kp}.
\end{eqnarray*}
Then
we get the decomposition given by (\ref{f305}),
using the   orthogonal transform and taking into account (\ref{b3012}).
\\
\\
\textbf{Example.} WHT basis images. The $2\times2$ orthogonal WHT matrix known also as Hadamard matrix
consists of plus 1 and minus 1
\begin{eqnarray}
&&H=\frac{1}{\sqrt{2}}
\begin{bmatrix}
1 &1 \\
1 &-1
\end{bmatrix}.
\end{eqnarray}
In optics this matrix describes  so called $50\%$ beam splitter,  a linear optical element often used
in experiments to split the beam into two parts.  Four basis images $a_{kp}$,  denoted as  tensor product of columns, have the following  form
\begin{eqnarray*}
&& a_{11}=\frac{1}{2}
\begin{bmatrix}
1 &1 \\
1 & 1
\end{bmatrix}, \ \ \ \ \ \ a_{12}=\frac{1}{2}
\begin{bmatrix}
1 &-1 \\
1 & -1
\end{bmatrix},
\\
&&\\
&&
a_{21}=\frac{1}{2}
\begin{bmatrix}
\ \ 1 &\ \ 1 \\
-1& -1
\end{bmatrix}, \ \
a_{22}=\frac{1}{2}
\begin{bmatrix}
\ \ 1 &-1 \\
-1 & \ \ 1
\end{bmatrix}.
\end{eqnarray*}
The determinant of every matrix equals to 0  and the matrices are non invertable.
The matrices  can be generated from a unit matrix by WHT:
\begin{eqnarray}
H: \ \ e_{11}=
\begin{bmatrix}
1 &0 \\
0 & 0
\end{bmatrix}
\leftrightarrows
\frac{1}{2}
\begin{bmatrix}
1 &1 \\
1 & 1
\end{bmatrix}
=a_{11}.
\end{eqnarray}
This equation illustrates  relations
between the basis images and the standard two-dimensional basis. But what is more interesting, the equation demonstrates
scattering  of  digital data (\ref{fa104}).
So,  a nonzero pixel of the unit matrix transforms  into a basis images of a matrix
with only nonzero  pixels.
\\
$\diamondsuit$
\\
As a  result, basis images can be produced by transformation of unit matrices.
 \\
 \\
\textbf{The quantum analogue.} The presented features allow us to consider  basis images as a representation of quantum operators.
These operators describe transitions of a physical system between its states or levels.
\\
\\
Let us assume that $\{ |k\rangle\}$ and $\{ |q\rangle\}$ are  two basis of a single particle Hilbert space
\begin{eqnarray*}
\label{3_003}
&&\sum_{k}|k\rangle\langle k|= 1,\\
&&\sum_{q}|q\rangle\langle q|= 1,
\end{eqnarray*}
 where
$k\in Z=\{1,2,\dots\}$, $q\in Q=\{x,y,\dots\}$.
Let the overlapping integrals  be real
\begin{eqnarray}
\label{f308}
&&\langle k| q\rangle^{*}=\langle q| k\rangle.
\end{eqnarray}
 Then we find a real matrix
$\tilde{U}_{qk}=\langle q| k\rangle$ that is orthogonal because $Z$ and $Q$ are complete basis.
\\
The following operator
\begin{eqnarray}
\label{3_0030}
&&|k\rangle\langle p|= \hat{a}_{kp},
\end{eqnarray}
where $k,p\in Z$, describes transition
 from the state or level
$|p\rangle$ into level $|k\rangle$.
If $k=p$, this operator is known as projection  operator.
\\
Using $Q$, the introduced operator (\ref{3_0030}) can be presents  as a real matrix
\begin{eqnarray*}
&&\langle x| \hat{a}_{kp}|y\rangle =a_{kp}(xy),
\end{eqnarray*}
where $x,y\in Q$. It is not difficult to understand,  that these matrices are
basis images,  considered above.
\\
Using  $Z$  we can present any single particle operator $\hat{F}$  as follows
\begin{eqnarray*}
&&\hat{F} =\sum_{kp} |k\rangle\langle p| \langle k|\hat{F}|p\rangle.
\end{eqnarray*}
Operator $F$ can be written as a matrix using $Q$ and (\ref{f308}), then the right part  of this equation takes the form (\ref{f305}).
As result we find  that some of  representations of  single particle operators can be considered as basis grayscale images.

\section{ Basis wavelet images}
Basis images can be generated by
DWT. In calculation the DWT techniques
do not use matrix methods
and  the basis wavelet images can be achieved by transform of standard basis.
\\
\\
\textbf{Wavelet coefficients.}
The DWT coefficients have a block structure due to  orthogonal matrix $U$.  In  case of single level transform this matrix consists of two parts  $L$ and  $H$ known as low and high frequency blocks. Let $G$ be a frequency representation of a $N\times N$ grayscale image $F=UGU^{T}$. Applying  the MATLAB notation, we write DWT as follows
\begin{eqnarray}
\label{b4011}
&&G=dwt(F)=
\begin{bmatrix}
cA & cH \\
cV & cD
\end{bmatrix},
\\
\nonumber
&&
\\
\nonumber
&&
F=idwt(cA,cH,cV,cD).
\end{eqnarray}
Here the introduced blocks $cA$, $cH$, $cV$ and $cD$ --- are  approximation coefficients, horizontal, vertical and diagonal details or $LL$, $LH$, $HL$ and  $HH$ frequency bands.
\\
The DWT coefficient matrix $G$ can be considered as a three-dimensional array $G=\{G_{kpz}\}$ of  size $N/2\times N/2\times 4$. Index $z=1,2,3,4$ labels  the $cA$, $cH$, $cV$ and $cD$ blocks, for example, $G_{kp1}=cA_{kp}$.
\\
\\
\textbf{Block structure of basis and basis images.}
To calculate basis images we use equation
(\ref{b3012})
\begin{eqnarray*}
&&a_{kp}= U e_{kp} U^{T}.
\end{eqnarray*}
According to  (\ref{b4011}) indexes $(k,p)$  belong to
one of the blocks $cA$, $cH$, $cV$ or $cD$.
Let $(k,p)\in cD$, so there is a set of basis items
\begin{eqnarray}
\label{f4014}
&&E_{(kpD)}=idwt(O,O,O,e_{kp}),
\end{eqnarray}
where $O$ --- is a $N/2\times N/2$ matrix of zeros. Here the upper indexes are  in brackets to label
number of the matrices instead of  indicating the pixel position.
In other words, we perform an orthogonal transformation of the unit block matrix
\begin{eqnarray*}
&&E_{(kpD)}\leftrightarrows
\begin{bmatrix}
O & O \\
O & e_{kp}
\end{bmatrix}.
\end{eqnarray*}
The total number of basis images of
$E_{D}=\{E_{(kpD)}\}$ is $N^{2}/4$, every  image is a $N\times N$ matrix.
\\
It is important to note that  the equation (\ref{b3012}) gives  solution by Matlab functions \texttt{dwt} and \texttt{idwt}. The reason is that in practice the DWT calculations are often  based on the filter function techniques \cite{Mallat}.
These techniques were developed for signal processing without referring to the orthogonal matrix $U$.  Usually wavelets are introduced numerically or by recurrent
equations so the  calculation of $U$ is a problem (except, for example, the Haar wavelet).
\\
 Using the block coefficients $cD$ and $E_{kpD}$
we can achieve an approximation
of original image
\begin{eqnarray*}
&&D=\sum_{kp}cD_{kp}E_{ (kpD)}.
\end{eqnarray*}
This image has diagonal details only.
\\
The wavelet coefficient structure results in basis of four blocks. The blocks
refer to $cA$, $cH$, $cV$ and $cD$ similarly to (\ref{f4014})
\begin{eqnarray*}
&&\Big\{\{E_{A}\}. \{E_{H}\}, \{E_{V}\}, \{E_{D}\}\Big\}.
\end{eqnarray*}
Every  block has $N^{2}/4$ basis  $N\times N$  images. As a result the representation over the wavelet basis images looks as follows
\begin{eqnarray*}
&&F=\sum_{kp}\Big( cA_{kp}E_{(kpA)}+
cH_{kp}E_{(kpH)}
+cV_{kp}E_{(kpV)}
+cD_{kp}E_{(kpD)}\Big).
\end{eqnarray*}
Indeed, the considered above function \texttt{dwt} can produce another basis. For this case in accordance with
(\ref{b4011}) every  basis images has a block structure
\begin{eqnarray*}
\label{0111}
&&J_{(xy)}=dwt(e_{xy})=
\begin{bmatrix}
J_{(xyA)} & J_{(xyH)} \\
J_{(xyV)} & J_{(xyD)}
\end{bmatrix},
\end{eqnarray*} 
\section{ A block matrix}
Basis images may be items of a matrix that can be  orthogonal.
\\
\\
\textbf{A matrix of basis images.} Consider a square $N\times N$ matrix, which elements are basis images
\begin{eqnarray}
\label{f5000}
\nonumber
&&b=\{b_{mn}\},\\
&&b_{mn}=a_{(nm)}.
\end{eqnarray}
Elements of $b$ do not commute.
The introduced matrix is a  four-dimensional  array, consisting of
$(N\times N)\times (N\times N)$
elements
\begin{eqnarray*}
&&K_{kpxy}=a_{kp}(x,y)=U_{xk}U_{yp}.
\end{eqnarray*}
\\
The matrix $b$ has the following important feature:
\begin{eqnarray}
\label{f5001}
&&bb=1.
\end{eqnarray}
So considering the matrix elements we find
\begin{eqnarray*}
&&(bb)_{mn}=\sum_{k}b_{mk}b_{kn}=\sum_{k}a_{km}a_{nk}
=\delta_{mn}\sum_{k}a_{kk}=\delta_{mn}.
\end{eqnarray*}
Indeed, matrix $\beta$, which elements are basis images, $\beta_{mn}=a_{mn}$, doesn't have  the property  given by
(\ref{f5001}). In this case  $\beta\beta=N\beta$.
\\
\\
\textbf{The biorthogonal decomposition.}
The equation (\ref{f5001}) tells that the matrix $b$ has rows  orthogonal to columns
\begin{eqnarray}
\label{f5002}
&&\langle b^{T}_{m}b_{n}\rangle =\delta_{mn}.
\end{eqnarray}
However, the rows are not orthogonal vectors themselves   and similarly to columns. An orthonormal basis is obtained from  rows and columns. The  basis is known to be  biorthogonal or biorthonormal  \cite{GanTmaher} and it can  be used to represent digital arrays.
\\
Let us consider a vector
$f=\{f_{k}\}$, $k=1,\dots, N$. Using (\ref{f5001}),  we find
\begin{eqnarray}
\label{f5007}
\nonumber
&&f=bbf=bg,\\
&&g=bf,
\end{eqnarray}
where the introduced vector
$g=\{g_{p}\}$, $p=1,\dots,N$ is a representation of $f$. To focus on the particular feature of
transform (\ref{f5007}), we introduce decomposition of vectors
$f$ and $g$ over columns of matrix $b$
\begin{eqnarray*}
&&f_=\sum_{k}b_{k} g_{k}, 
\end{eqnarray*}
In contrast to orthogonal transform, the coefficients $g_{k}$ are denoted by rows but not by columns
\begin{eqnarray*}
&&g_{k}= \langle b_{k}^{T},f\rangle.
\end{eqnarray*}
That is a biorthogonal decomposition.
\\
The biorthogonal decompositions are applied  in the wavelet field. So, to perform      \texttt{dwt} (\ref{b4011}) and inverse transform \texttt{idwt},  we need two  different wavelets. An example is the Cohen-Daubechies-Feauveau wavelet  or biorthogonal  9/7 wavelet  that is used in JPEG 2000.
\\
\\
\textbf{Orthogonality.} Is the  matrix $b$ orthogonal?
The answer is not clear  because  $b$ is a four-dimensional array.  However, we can refer to the array primitives and consider  rows and columns consisting of the rows and columns of the basis images. Let $r_{k}$ be a block row. It has items $b_{k1}, b_{k2},\dots,b_{kN}$  of basis images
$a_{1k},a_{2k},\dots,a_{Nk}$. Selecting a row $x$ of every basis image, we get a row $r_{kx}$. This  may be done for a column as well. Introduced  rows and columns   will
be  orthonormal vectors.  This is a reason to consider  the block matrix $b$  as  an orthogonal matrix.
\\
Indeed,  this result follows
from the definition of the transposing operation. In  case of  block matrix $Z$
it can be presented as follows
\begin{eqnarray*}
&&Z^{T}=
\begin{bmatrix}
Z_{11}&Z_{12}\\
Z_{21}& Z_{22}
\end{bmatrix}^{T}
=
\begin{bmatrix}
Z_{11}^{T}&Z_{21}^{T}\\
Z_{12}^{T}& Z_{22}^{T}
\end{bmatrix}.
\end{eqnarray*}

\section{ The block based representation}
The orthogonal matrix  of  basis images
provides  a block based representation of  multi-dimensional   arrays.
\\
\\
\textbf{Representation.} The block based representation follows from the equations (\ref{f5007}), if they are  written in the matrix form
\begin{eqnarray}
\label{f6001}
\nonumber
&&f_{k}=\sum_{k}a_{pk}g_{p},\\
&& g_{p}=\sum_{k}a_{kp}f_{k}.
\end{eqnarray}
Here $f$ and $g$ are two block vectors
\begin{eqnarray*}
&&f=( f_{1}, f_{2},\dots, f_{N}),\\
&&g=( g_{1}, g_{2},\dots, g_{N}).
\end{eqnarray*}
items of which may be chosen as  vectors, matrices etc.
\\
Let us assume   that  $f_{k}=\{f_{k}(x,y)\}$ and $g_{p}=\{g_{p}(x,y)\}$ are   $N\times Q$ matrices. Restrictions on $Q$ will be established later.
For this case the equations (\ref{f6001}) take the following form
\begin{eqnarray}
\label{f6002}
\nonumber
&&f_{k}(x,y)=\sum_{p,z}a_{pk}(x,z)g_{p}(z,y),\\
&&g_{p}(x,y)=\sum_{k,z}a_{kp}(x,z)f_{k}(z,y).
\end{eqnarray}
It is important to notice that index $y$ plays minor role in these equations   and  from  the formal point of view it is unnecessary.
It means  that  $f_{k}$ and $g_{p}$ have to be not less than one-dimensional   arrays.  Then for  considered  matrices we find  the following   condition $Q\geq 1$.
\\
The unnecessary index indicates  that there is more  space to  which matrices $a_{kp}$ do not belong. From the physical point of view  we have two systems, for example, atoms and light. Both systems are described by its observations that  can be represented by matrices  that, however,  affect   its Hilbert spaces. To describe elements of different spaces, e.g.  two matrices  $A$ and $B$ a  tensor product is introduced $A\otimes B$.

\section{ Non separability and scattering}
The block based representation leads to new features  of data scattering and has a quantum analogue.
\\
\\
\textbf{Correlation.}
Formally, the block based representation (\ref{f6002}) looks as one-dimensional transform (\ref{f103}) and we find
properties given by (\ref{f104})
for data scattering. However,  due to large number of degrees of freedom, scattering obtains new features.
\\
\\
Let us assume that  both arrays $f_{k}$ and $g_{p}$ are  block matrices consisting of other matrices. They are four-dimensional  arrays  that
we specify by four indexes $(x,y,\alpha,\beta)$. Let, in contrast to $g_{p}$, the array $f_{k}$ be dependant  on the last pair of indexes only
\begin{eqnarray*}
&&f_{k}=\delta_{xy}\psi_{k}(\alpha,\beta),\\
&&g_{p}=g_{p}(x,y,\alpha,\beta).
\end{eqnarray*}
Under these conditions the equations (\ref{f6002}) take the following form
\begin{eqnarray}
\label{f7001}
\nonumber
&&1\otimes \psi_{k}=\sum_{p}(a_{pk}\otimes 1)g_{p},\\
&&g_{p}=\sum_{k}a_{kp}\otimes \psi_{k},
\end{eqnarray}
where $1\otimes \psi_{k}=f_{k}$.
\\
 An important fact   follows that
  the  array $g_{p}$ is  non-separable.   We will use the term separable as
divisibility, when the variables are factorized. For example, the function
$F(x,y)=\cos(x)\cos y$ is separable over $x$ and $y$ and the function
$ \Phi(x,y)=\cos(x+y)$  is not.  In our case we focus on two pairs of variables, a pair  $x,y$, that describe basis images
$a_{pk}(x,y)$, and pair $\alpha,\beta$. From this point  of view, the array  $f_{k}$ is separable in contrast to $g_{p}$, that is  a non-separable array, because it is a sum of products
\begin{eqnarray}
&&g_{p}(x,y,\alpha,\beta)=a_{1p}(x,y) \psi_{1}(\alpha,\beta)+
a_{2p}(x,y) \psi_{2}(\alpha,\beta) +\dots
\end{eqnarray}
Non separability is a kind of correlation. Now this is a correlation between the matrices from different spaces, the basis images and  the matrices $\psi_{k}$.
\\
\\
\textbf{Scattering.}
Due to the property
of the scalar product of basis matrices
(\ref{b3000}),
we find that
\begin{eqnarray*}
&&\sum_{xy}a_{kp}(x,y)g_{p}(x,y,\alpha,\beta)=\psi_{k}(\alpha,\beta)
\end{eqnarray*}
or
\begin{eqnarray}
\label{f6006}
&&\langle a_{kp}, g_{p}\rangle=\psi_{k}.
\end{eqnarray}
For data scattering this result tells us the following.
The component $\psi_{k}$ scatters into every $g_{p}$ with its weight $a_{kp}$ and it may be established from every item $g_{p}$
\begin{eqnarray*}
&&\psi_{k}\to \{g_{1}, g_{2},\dots,\}, \\
&&\psi_{k} \leftarrow g_{p}.
\end{eqnarray*}
This is a new property and  it is usually impossible.
The property arises from non separability produced by orthogonal transform of  the block matrix $b$. The transform results in correlation between
the set of basis images and the input matrices.
\\
\\
 \textbf{A quantum analogue.}
The  block-based representation can be introduced for a three particle  quantum system.
\\
Let us consider a three particle operator given by
\begin{eqnarray*}
&& \hat{c}= \sum_{k,p} |k\rangle  \langle p|\otimes
|p \rangle \langle k|\otimes 1,
\end{eqnarray*}
where $k,p \in Z=\{1,2,\dots\}$ and $\{|k \rangle \}$ is a single particle basis. The operator $\hat{c}$ is Hermitian and unitary
\begin{eqnarray*}
&& \hat{c}=\hat{c}^{\dagger},\\
&& \hat{c}\hat{c}=1.
\end{eqnarray*}
Let us note that  two particle operator
\begin{eqnarray*}
&& \hat{b}=\sum_{k,p} |k\rangle \langle p|\otimes |p \rangle \langle k|=\hat{b}^{\dagger}
\end{eqnarray*}
is a quantum analogue of the matrix
$b$, given by (\ref{f5000}).
\\
Let us introduce   three particle operators $\hat{f}$ and $\hat{g}$ that are equal up to orthogonal  transform
 given by $\hat{c}$
\begin{eqnarray*}
&&\hat{f}=\hat{c}\hat{g}, \\
&& \hat{g}=\hat{c}\hat{f}.
\end{eqnarray*}
These equations can be written in a block form. Introducing  the matrix elements over particle 1  for operators
$\hat{f}$ and $\hat{g}$ way we get two operators
of particle 2 and 3, which we  denote as
\begin{eqnarray*}
&&_{1}\langle k|\hat{f}|m\rangle_{1} =
\hat{f_{k}},\\
&&_{1}\langle p|\hat{g}|m\rangle_{1} =\hat{g_{p}},
\end{eqnarray*}
where $k,p, m \in Z$. Then we have the block representation
 \begin{eqnarray*}
&&\hat{f}_{k}=\sum_{p}(\hat{a}_{pk}\otimes 1)\hat{g}_{p},\\
&&\hat{g}_{p}=\sum_{k}(\hat{a}_{kp}\otimes 1)\hat{f}_{k}.
\end{eqnarray*}
Let  $\hat{f}_{k}$  be the operator of particle 3 only,
$\hat{f}_{k}=1\otimes \hat{\psi}_{k}$,
then  we find  $\hat{f}_{k}$ being  a two particle non-separable operator
\begin{eqnarray*}
&&\hat{g_{p}}=\hat{a_{1p}}\otimes \hat{\psi}_{1}+
\hat{a_{2p}}\otimes \hat{\psi}_{2}\dots.
\end{eqnarray*}
This equation  is a quantum  analogue  of (\ref{f6006}) found  for digital data scattering.  It is obvious, that
\begin{eqnarray*}
  &&Sp_{2}\{\hat{a_{kp}} \hat{g_{p}}\}=\hat{\psi_{k}},
  \end{eqnarray*}
where the average refers to particle 2.

\section{ A steganographic scheme}
The block-based representation may be useful for frequency domain steganographic technique.
\\
\\
\textbf{Scheme.} Let the  digital data $f$
be images in a spatial domain and $g$ be its representation in a frequency domain.  Any standard frequency embedding scheme has the following  steps.
\begin{itemize}
\item Transform data into the frequency domain $f\to g$ and embed a message $M$ using an algorithm
$g\to g_{M}= emb(g, M, K)$, where $K$ is a set of parameters with a possible secrete key.
\item Transform data into the spatial domain
$g_{M}\to f_{M}$ and send it to a receiver via  the communication channel.
\item Extract the embedded message using  detection algorithm $f_{M}\to M=det(f_{M},K)$.
\end{itemize}
The scheme includes  transformations
\begin{eqnarray*}
&&f\to g \to g_{M} \to f_{M} \to g_{M}\to M.
\end{eqnarray*}
Indeed, the transform $g_{M}\to f_{M}$ can scatter the embedded data among the spatial domain. Scattering may result in more robust of hidden data
to degradation due to various transformations.
An example is a JPEG lossy compression, that stores image in a graphical format. By decreasing the image redundancy, the lossy compression introduces changes into embedded data that exploits  the redundancy.
So, there is a trade  between the compression and  the quality of the extracted information. The higher  the    compression level is,  the worse the quality is.
\\
\\
\textbf{Data scattering in the spatial  domain.}
Let us consider data scattering in the block based representation
({\ref{f6001}), assuming $k,p=1,2$. Let a message be embedded into $g_{2}\to g_{2M}$. Then two spatial items will be changed
\begin{eqnarray}
\label{f8001}
\nonumber
&&f_{1}\to f_{1M}=a_{11}f_{1}+a_{21}g_{2M},\\
&&f_{2}\to f_{2M}=a_{11}f_{2}+a_{22}g_{2M}.
\end{eqnarray}
To extract the  message, we need
$g_{2M}$ or two items $f_{1M}$ and $f_{2M}$
\begin{eqnarray*}
\label{f8002}
\nonumber
&&g_{2M}= a_{12}f_{1M}+a_{22}f_{2M}.
\end{eqnarray*}
The equation is  a basis  for the detection algorithm
\begin{eqnarray}
\label{f8002}
\nonumber
&& det(f_{1M}, f_{2M}, K)\to M.
\end{eqnarray}
Data scattering means that all spatial items were changed after embedding and all  items are required for detection. Any frequency domain watermarking technique has  these properties
regardless of whether it use the block based representation or not.
However, the representation  leads to  new features appearance.
\\
\\
\textbf{Embedding.}
Let us assume that both vectors $f$ and  $g$ have two components. Also  $f_{k}=1\otimes \psi_{k}$, where $\psi_{k}$ is an image in the spatial domain, $k=1,2$. In accordance to  (\ref{f7001}), the frequency representation $g$ consists of pair of
four-dimensional  arrays
\begin{eqnarray*}
\label{f8002a}
\nonumber
&&  g_{1}=a_{11}\otimes \psi_{1}+a_{21}\otimes \psi_{2},\\
&& g_{2}=a_{12}\otimes \psi_{1}+a_{22}\otimes \psi_{2}.
\end{eqnarray*}
Let the embedding algorithm  replace  $\psi_{k}$ with  messages
\begin{eqnarray}
\label{f8002}
\nonumber
&& g_{1}\to g_{1M}=a_{11}\otimes M_{1}+a_{21}\otimes M_{2},\\
&& g_{2}\to g_{2M}=a_{12}\otimes M_{3}+a_{22}\otimes M_{4},
\end{eqnarray}
where four matrices $M_{1},\dots, M_{4}$ are introduced messages.
The main feature of this algorithm is   store the ability  to  the structure of the array  that holds   a set  of tensor products including basis images. In the spatial domain we have
\begin{eqnarray*}
&& f_{1M}=a_{11}\otimes M_{1}+a_{22}\otimes M_{3},\\
&& f_{2M}=a_{11}\otimes M_{2}+a_{22}\otimes M_{4}.
\end{eqnarray*}
This allows us to exploit the equation (\ref{f6006}) for detection. Then the embedded messages can be extracted, if the  component
$f_{1M}$ or the component $f_{2M}$ is  given
\begin{eqnarray*}
&& f_{1M}\to M_{1}= \langle a_{11}, f_{1M}\rangle.
\end{eqnarray*}
For this case the detection algorithm works as follows
\begin{eqnarray*}
&& det(f_{1M}, K)\to M_{1}, M_{3},\\
&& det(f_{2M}, K)\to M_{2}, M_{4}.
\end{eqnarray*}
Let us note that it differs from the standard algorithm (\ref{f8002}) that needs two spatial items instead of one.
\\
Moreover there is a difference  between this  and   the spatial domain embedding.
We assume  that message is  embedded into  spatial components
$\psi_{k}\to \psi_{kM}$, $k=1,2$. It is clear that  two messages can be embedded only. In the frequency domain there are four messages  that may be embedded.  But what is more important, these four messages can be distinguished. This fact plays a key role in detection and arises from coupling the messages and basis images to be orthogonal and hence to be well distinguished.
\\
\\
Taking into account the considered quantum analogues, we admit   that the presented scheme can be extended to quantum mechanic fields.

\section {Conclusions}
\begin{enumerate}
\item Orthogonal transform provides decomposition over basis items or basis images that have a simple
quantum analogue. So, they are a representation of single particle operators that describe transitions
of a particle between its states.
\item Grayscale, color and wavelet basis images can be introduced for decomposition of two- and three-dimensional arrays.
\item Basis images can be achieved by orthogonal transform of a standard basis that is a set of unit vectors,
unit matrices and etc. This fact illustrates digital data scattering, a process of redistributing pixel
energy.
\item Due to scattering, energy can be concentrated in small amount of items or, in contrast, be spread.
Both cases are interesting for applications. For example, in lossy compression scattering allows to
extract the image redundancy, in watermarking it can increase the robustness of a watermark.
\item A block matrix of basis images may be orthogonal and suitable for transformation of multi-dimensional
arrays. Different degrees of freedom can be correlated by this transform and non separable
arrays can be produced. As a result, in this way, new features of scattering appears. These
features may be used whole executing detection algorithms in frequency domain watermarking.
\end{enumerate}

\end{document}